\documentclass[twocolumn,10pt,prl]{revtex4}

\def\mysections#1{{\bf #1.} }

\usepackage[dvips]{graphicx}
\usepackage{epsfig,amsmath,amssymb,verbatim,mathrsfs,array,layout,textcomp,amssymb,latexsym, slashed,graphicx,bbm}

\usepackage{graphicx}
\usepackage{amsmath}
\usepackage{amssymb}
\usepackage{appendix}
\usepackage{verbatim,mathrsfs,array,layout,textcomp,latexsym, slashed}

\usepackage{amsmath, amssymb, graphicx, rotating}
\usepackage{hyperref}
\usepackage[usenames,dvipsnames]{color}

\newcommand{\beq}{\begin{eqnarray}}
\newcommand{\eeq}{\end{eqnarray}}

\def\beqa{\begin{eqnarray}}
\def\eeqa{\end{eqnarray}}
\newcommand{\no}{\nonumber}
\newcommand{\bv}{\left(\begin{array}{c}}
\newcommand{\ev}{\end{array}\right)}
\newcommand{\bmtwo}{\left(\begin{array}{cc}}
\newcommand{\bmthree}{\left(\begin{array}{ccc}}
\newcommand{\emn}{\end{array}\right)}
\newcommand{\bmtwoc}{\left\{\begin{array}{cc}}
\newcommand{\bmthreec}{\left\{\begin{array}{ccc}}
\newcommand{\emnc}{\end{array}\right\}}
\newcommand{\ba}{\begin{array}}
\newcommand{\ea}{\end{array}}

\newcommand{\LL}{\mathcal{L}}

\newcommand{\Tr}{{\text{Tr }}}

\newcommand{\units}[1]{\mathrm{\; #1}}

\def\lsim{\mathrel{\rlap{\lower4pt\hbox{\hskip1pt$\sim$}}
     \raise1pt\hbox{$<$}}}         
\def\gsim{\mathrel{\rlap{\lower4pt\hbox{\hskip1pt$\sim$}}
     \raise1pt\hbox{$>$}}}         

\begin{document}

\font\mini=cmr10 at 0.8pt

\title{
The SIMPlest Miracle
}

\author{Yonit Hochberg${}^{1,2}$}\email{yonit.hochberg@berkeley.edu}
\author{Eric Kuflik${}^3$}\email{kuflik@cornell.edu}
\author{Hitoshi Murayama${}^{1,2,4}$}\email{hitoshi@berkeley.edu, hitoshi.murayama@ipmu.jp}
\author{Tomer Volansky${}^5$}\email{tomerv@post.tau.ac.il}
\author{Jay G. Wacker${}^{6,7}$}\email{jgwacker@stanford.edu}
\affiliation{${}^1$Ernest Orlando Lawrence Berkeley National Laboratory, University of California, Berkeley, CA 94720, USA}
\affiliation{${}^2$Department of Physics, University of California, Berkeley, CA 94720, USA}
\affiliation{${}^3$Department of Physics, LEPP, Cornell University, Ithaca NY 14853, USA}
\affiliation{${}^4$Kavli Institute for the Physics and Mathematics of the
  Universe (WPI), Todai Institutes for Advanced Study, University of Tokyo,
  Kashiwa 277-8583, Japan}
\affiliation{${}^5$Department of Physics, Tel Aviv University, Tel Aviv, Israel}
\affiliation{${}^6$Quora,  Mountain View, CA  94041 USA}
\affiliation{${}^7$Stanford Institute for Theoretical Physics, Stanford University,  Stanford, CA  94305 USA}

\begin{abstract}
It has recently been proposed that dark matter could be a thermal relic of $3\to2$ scatterings in a strongly coupled hidden sector. We present explicit classes of strongly coupled gauge theories that admit this behavior. These are QCD-like theories of dynamical chiral symmetry breaking, where the pions play the role of dark matter. The number-changing $3\to2$ process, which sets the dark matter relic abundance, arises from the Wess-Zumino-Witten term. The theories give an explicit relationship between the $3\to2$ annihilation rate and the $2\to2$ self-scattering rate, which alters predictions for structure formation. This is a simple calculable realization of the strongly-interacting-massive-particle (SIMP) mechanism.
\end{abstract}

\maketitle

\section{Introduction}
The majority of the matter content of our universe is in the form of dark matter (DM).  An appealing explanation for its measured abundance
 is that it is a thermal relic of the early universe. The most well-studied thermal scenario is that of a weakly-interacting-massive-particle (WIMP), whose relic abundance is set by $2\to 2$ annihilations, typically into Standard Model (SM) particles. This mechanism predicts dark matter mass of order the weak scale for coupling of order the weak coupling.

Ref.~\cite{Hochberg:2014dra} proposed a new paradigm for achieving thermal relic dark matter. The requisite features of the mechanism
are the following:
\begin{itemize}
  \item The dark matter relic abundance is set thermally by the freeze-out of a $3\to 2$ process that reduces the number of dark matter particles within the dark sector.
    \item At the time of freeze-out, dark matter is in thermal equilibrium with the SM.
\end{itemize}
This setup, termed the strongly-interacting-massive-particle (SIMP) mechanism,
robustly predicts light dark matter with mass in the MeV to GeV range, with strong self-interactions.
Annihilations into SM particles are subdominant during freeze-out, but DM scattering off the SM bath is fast enough to maintain kinetic equilibrium between the dark and visible sectors.
 The strongly interacting hidden sector is expected to contribute to DM self-scattering cross-sections that are relevant for structure formation.

In what follows, we find explicit strongly coupled realizations for the hidden sector that admit the $3\to2$ process of the SIMP mechanism. Explicit viable realizations for the mediation mechanism between the dark and visible sectors exist, and will be presented in detail in a forthcoming publication~\cite{future}.

\section{The SIMPlest realization}\label{sec:WZW}

In Ref.~\cite{Hochberg:2014dra}, a weakly coupled toy model which incorporates the SIMP mechanism and leads to stable dark matter was presented.
Here we present three classes of strongly coupled gauge theories that realize the SIMP mechanism. The basic idea is as follows. We use the well-known 5-point interaction term present in theories of chiral symmetry breaking, first discovered by Wess and Zumino~\cite{Wess:1971yu} and later studied by Witten~\cite{Witten:1983tw, Witten:1983tx}, as the source of the $3\to2$ interactions. This term in massless QCD describes the low energy limit of two kaons annihilating into three pions. In a given theory, the existence of the Wess-Zumino-Witten (WZW) term is dictated by a topological condition on the symmetry-breaking pattern; for coset spaces with non-trivial fifth homotopy groups, the WZW term is non-vanishing.
This 5-point interaction then generates the $3\to2$ freeze-out process. In what follows, we demonstrate this explicitly.

We first consider and review an Sp($N_c$) gauge theory with $2 N_f$ Weyl fermions in the fundamental $N_c$-dimensional representation (with the number of colors $N_c$ even). In the massless limit the UV description takes a simple form
\begin{equation}\label{eq:L}
  {\cal L}_{\rm SIMP}
  = -\frac{1}{4} F_{\mu\nu}^a F^{\mu\nu a} + \bar{q}_i i {\not \!\! D}
  q_i \,, \quad i=1,\ldots 2N_f\,,
\end{equation}
which admits a global SU($2 N_f$) symmetry among the Weyl fermions $q_i$.
It is believed that this model, for moderately small $N_f$ in the asymptotically free range, leads to chiral symmetry breaking with
the order parameter
\begin{align}\label{eq:order}
\langle q_i q_j \rangle = \mu^3 J_{ij},
\end{align}
where $\mu$ is of mass dimension one and $J = i\sigma_2 \otimes \mathbbm{1}_{N_f}$ is a $2 N_f\times 2 N_f$ anti-symmetric matrix
that preserves an Sp($2N_f$) subgroup of the SU($2 N_f$) flavor symmetry~\cite{Peskin:1980gc,Preskill:1980mz,Witten:1983tx,Kosower:1984aw}. For $N_f\geq 2$, the topological condition is met,
\beq
\pi_5({\rm SU}(2N_f)/{\rm Sp}(2N_f))={\mathbb Z}\,,\quad N_f\geq 2\,,
\eeq
and the WZW term is non-vanishing. The coset space ${\rm SU}(2 N_f)/{\rm Sp}(2N_f)$ is a symmetric space and is parameterized by
$N_\pi=2N_f^2-N_f-1$ pion fields, $\pi^a$, corresponding to the broken generators $T^a$, with $a=1,\ldots N_\pi$. The pions furnish a rank-two anti-symmetric tensor representation of the unbroken Sp$(2N_f)$, and are stable. Assuming the pions are the lightest states in the theory, dark matter is comprised of these $N_\pi$ pions.

A simple parametrization is found by performing a transformation on the vacuum and promoting the transformation parameters to fields,
\begin{align}\label{eq:vev}
\langle q q \rangle = \mu^3 J \rightarrow \mu^3 V J V^T \equiv \mu^3 \Sigma\,,
\end{align}
where $V=\exp(i \pi/ f_\pi)$ and $f_\pi$ is the decay constant. Since the broken generators obey $\pi J - J\pi^T =0$ with $\pi=\pi^a T^a$ and ${\rm Tr}(T^a T^b)=2\delta^{ab}$, we have
\begin{eqnarray}\label{eq:sigma}
\Sigma = \exp( 2i \pi/f_\pi) J \,.
\end{eqnarray}
A minimal realization of the $3\to2$ mechanism is an Sp(2)~$\simeq$~SU(2) gauge theory with $N_f=2$ flavors. Dark matter is comprised of 5 pions that transform as a 5-plet under the preserved Sp(4) flavor symmetry.  The coset space of SU($4$)/Sp($4$) = SO($6$)/SO($5$) is then topologically an $S^5$. (See {\it e.g.} Refs.~\cite{Hands:1999md,Kogut:1999iv,Hands:2000ei,Aloisio:2000if,Kogut:2001na,Hands:2000yh,Aloisio:2000rb,Kogut:2001if,Kogut:2002cm,Kogut:2003ju,Nishida:2003uj,Lombardo:2008vc}  for lattice work on low-lying spectra in the minimal Sp(2) with quarks in the fundamental representation, and Refs.~\cite{Ryttov:2008xe,Lewis:2011zb,Detmold:2014qqa,Detmold:2014kba} for dark-matter examples.)

The relevant pion Lagrangian receives contributions from several terms.
The canonically normalized kinetic term yields kinetic and 4-point interactions for the pions,
\beq\label{eq:kin}
\LL_{\text{kin}}&=& \frac{f_\pi^2}{16} \Tr \partial_\mu \Sigma\; \partial^\mu \Sigma^\dagger\\
&=& \frac{1}{4}\Tr \partial_{\mu} \pi \partial^{\mu }\pi \no\\
&&-\frac{1}{6 f_\pi^2} \Tr\left(\pi^2 \partial^{\mu} \pi \partial_{\mu}  \pi  -\pi \partial^{\mu} \pi \pi \partial_{\mu} \pi \right)+{\cal O}(\pi^6/f_\pi^4)\,,\no
\eeq
where in our normalization, ${\rm Tr}(\pi^2)=2\pi^a\pi^a$.
The Wess-Zumino-Witten term~\cite{Wess:1971yu,Witten:1983tw} yields 5-point pion interactions. It can be written as an integral on the boundary of a five-dimensional disk, identified with our four-dimensional spacetime,
\begin{equation}
  {\cal S}_{\rm WZW}
  = \frac{-i  N_c}{240\pi^2} \int \Tr (\Sigma^\dagger d \Sigma)^5\,.
\end{equation}
To leading order in pion fields,
\begin{equation}\label{eq:wzw}
  {\cal L}_{\rm WZW}
  = \frac{2N_c}{15 \pi^2 f_\pi^5}\epsilon^{\mu\nu\rho\sigma} \Tr\!\!\left[\pi \partial_\mu
  \pi \partial_\nu \pi \partial_\rho \pi \partial_\sigma \pi \right],
\end{equation}
which is responsible for the required $3\rightarrow 2$ annihilation process.
Finally, an Sp($2N_f$)-preserving mass term can be written for the quarks:
\begin{align}\label{eq:mass}
 {\cal L}_{\rm mass}=- \frac{1}{2}
M^{ij} q_i q_j + c.c.,\quad  M^{ij} = m_Q\; J^{ij}\,.
\end{align}
The pions are then pseudo-Goldstone bosons of the broken symmetry and acquire a mass, as well as contact interactions:
\beq\label{eq:eff}
  \Delta{\cal L}_{\text{eff}} &=& -\frac{1}{2} m_Q \mu^3 {\rm Tr} J \Sigma + c.c.\\
 &=&-\frac{m_\pi^2}{4} \Tr\!  \pi^2 + \frac{m_\pi^2}{12 f_\pi^2}  \Tr\! \pi^4\,+{\cal O}(\pi^6/f_\pi^4)\,,\no
\eeq
where
\begin{align}\label{eq:pimass}
m_\pi^2  = 8\frac{  m_Q \mu^3}{ f_\pi^2}\,.
\end{align}
Combining all the above we arrive at the relevant pion Lagrangian,
\beq\label{eq:Lpion}
{\cal L}_\pi &=&  {\cal L}_{\rm kin} + \Delta {\cal L}_{\rm eff} + {\cal L}_{\rm WZW} \\
 &=& \frac{1}{4}\Tr \partial_{\mu} \pi \partial^{\mu }\pi -\frac{m_\pi^2}{4} \Tr\!  \pi^2 +\frac{m_\pi^2}{12 f_\pi^2}  \Tr\! \pi^4 \no\\ &&-\frac{1}{6 f_\pi^2} \Tr\left(\pi^2 \partial^{\mu} \pi \partial_{\mu}  \pi  -\pi \partial^{\mu} \pi \pi \partial_{\mu} \pi \right) \no\\
 && +\frac{2N_c}{15 \pi^2 f_\pi^5}\epsilon^{\mu\nu\rho\sigma} \Tr\!\!\left[\pi \partial_\mu
  \pi \partial_\nu \pi \partial_\rho \pi \partial_\sigma \pi \right]+{\cal O}(\pi^6)\,.\no
\eeq

There can also be ${\cal O}(\pi^4)$ terms with four derivatives and higher. These contribute to four-pion self-scattering with a naive-dimensional-analysis~\cite{Georgi:1992dw} suppression of at least ${\cal O}(m_\pi^2/\Lambda^2)$, where $\Lambda=2\pi f_\pi$, compared to those we keep. The ${\cal O}(\pi^5)$ terms with four derivatives that we use are the leading 5-point pion interactions of the theory.

The same principle presented above to construct strongly coupled models, that admit $3\to2$ interactions and realize the SIMP mechanism, is generalizable to other gauge and flavor symmetries. For instance, one can consider a generalized QCD-like theory with an SU($N_c$) gauge group and $N_f$ Dirac-fermions in the fundamental representation. The global flavor symmetry of the theory is SU($N_f$)$\times$SU($N_f$), which upon chiral symmetry breaking preserves an SU($N_f$) subgroup. Similarly, an O($N_c$) gauge group with $N_f$ fermions in the vector representation exhibits an SU($N_f$) flavor symmetry, which breaks to SO($N_f$) once chiral symmetry breaking occurs. The topological condition on the coset space in each of these cases,
\beq
\pi_5({\rm SU}(N_f))&=&{\mathbb Z}\,,\quad N_f\geq3\,, \no \\
\pi_5({\rm SU}(N_f)/{\rm SO}(N_f))&=&{\mathbb Z}\,, \quad N_f\geq3 \,,
\eeq
admits the WZW term, Eq.~\eqref{eq:wzw}, and $3\to2$ pion interactions are present.
 The relevant pion Lagrangian terms in each of these cases are readily obtained by replacing $J$, the Sp($2N_f$) invariant, in Eqs.~\eqref{eq:order},~\eqref{eq:vev},~\eqref{eq:sigma},~\eqref{eq:mass} and~\eqref{eq:eff}, by $\mathbbm{1}$, which is the SU($N_f$) and SO($N_f$) invariant.
 Additionally, for the SU($N_c$) case, the quark bilinear in Eqs.~\eqref{eq:order} and~\eqref{eq:vev} is understood as $q \bar q$, with appropriate modification to the transformation in Eq.~\eqref{eq:vev}.
  Reading off the pion interactions accordingly, the Lagrangian of interest remains Eq.~\eqref{eq:Lpion}, with a change of sign in the relation Eq.~\eqref{eq:pimass}.

In what follows, we explicitly study the self-interactions of the pions in the three classes of gauge theories presented, namely Sp($N_c$), SU($N_c$) and O($N_c$). These theories yield qualitatively similar results, with some small quantitative differences described below. An exception arises due to the fact that, in contrast to the Sp($N_c$) class of models, baryons can exist in the SU($N_c$) and O($N_c$) theories.
The omission of baryons and other resonances is justified as long as these states are much heavier than the pions.

\section{Results}

\begin{figure*}[t!]
\begin{center}
\includegraphics[width=1\textwidth]{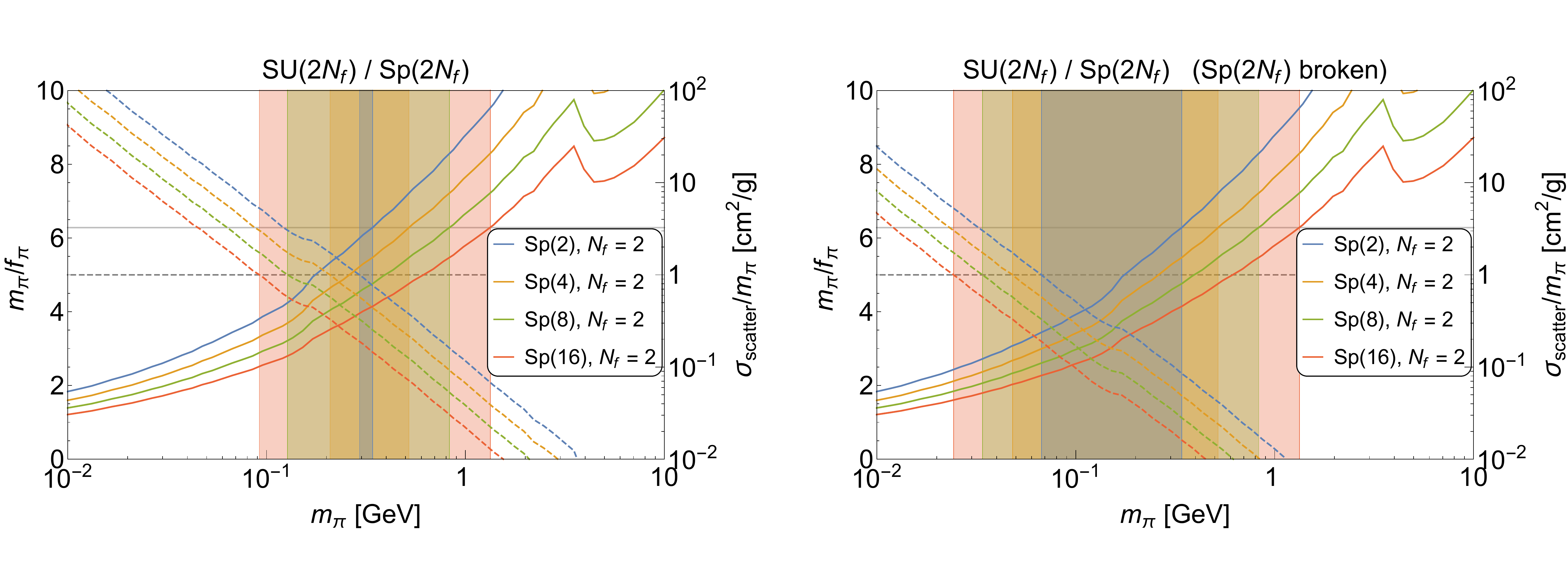}
\caption{
\label{fig:3to2sol}
{\bf Solid curves}: The solution to the Boltzmann equation of the $3\to2$ system, yielding the measured dark matter relic abundance for the pions, $m_\pi/f_\pi$,   as a function of the pion mass ({\bf left axis}). {\bf Dashed curves}: The self-scattering cross section along the solution to the Boltzmann equation, $\sigma_{\rm scatter}/m_\pi$, as a function of pion mass ({\bf right axis}). All curves are for selected values of $N_c$ and $N_f$, for an Sp($N_c$) gauge group with a conserved ({\bf left panel}) or broken ({\bf right panel}) Sp($2 N_f$) flavor symmetry.
The solid horizontal line depicts the perturbative limit of $m_\pi/f_\pi\lsim 2\pi$, providing a rough upper limit on the pion mass; the dashed horizontal line depicts the bullet-cluster and halo shape constraints on the self-scattering cross section, Eq.~\eqref{eq:bullet}, placing a rough lower limit on the pion mass. Each shaded region depicts the resulting approximate range for $m_\pi$ for the corresponding symmetry structure.
}
\end{center}
\end{figure*}

The WZW term in Eq.~\eqref{eq:wzw} induces the number-changing process that is responsible for the freeze-out of the pions. The Boltzmann equation governing the pion system is given by~\cite{Hochberg:2014dra}
\beq\label{eq:bol}
\begin{array}{rcl}
\dot{n}_{\pi} + 3 H {n}_{\pi}
&=&  -(n_{\pi}^3 - n_{\pi}^2 n_{\pi}^{\rm eq}) {\langle \sigma v^2 \rangle}_{3 \to 2}\,,
\end{array}\eeq
where $H$ is the Hubble constant, $n_{\pi}$ is the total pion number density, $n_\pi^{\rm eq}$ is their equilibrium number density, and ${\langle \sigma v^2 \rangle}_{3 \to 2}$ is the thermally averaged $3\to2$ cross section averaged over initial and final state pions, computed from the WZW term in Eq.~\eqref{eq:wzw}. As discussed in Ref.~\cite{Hochberg:2014dra},
within the SIMP mechanism,  $2\to2$ annihilations into SM particles may be neglected above.
Interactions maintaining thermal equilibrium with the thermal bath ensure that the dark matter has the same temperature as the photons.

In the left panel of Fig.~\ref{fig:3to2sol} we plot the results of solving the Boltzmann equation, Eq.~\eqref{eq:bol}, yielding the measured dark matter relic abundance, for an Sp($N_c$) gauge group for several values of $N_c$ and $N_f$ (solid curves). The results for the other gauge groups are very similar and are given in the left panel of Fig.~\ref{fig:others} in the Appendix. In our convention, an approximate perturbative limit of $m_\pi/f_\pi\lsim 2\pi$ (depicted by the horizontal solid line) can be set, placing a rough upper limit on the mass of the pions. Note that in the SM, $m_K/f_\pi\simeq 2.7$. The upper limit relaxes as $N_c$ increases and $N_f$ decreases. The reason is that the thermally averaged cross section at freeze-out arising from Eq.~\eqref{eq:wzw} is
\beq\label{eq:xsec32}
\langle \sigma v^2\rangle_{3\to2} = \frac{5\sqrt{5}}{2\pi^5 x_f^2}\;\frac{N_c^2 m_\pi^5}{f_\pi^{10}}\; \frac{t^2}{N_\pi^3}\,,
\eeq
where $x_f=m_\pi/T_f \simeq 20$ with $T_f$ the temperature at freeze-out.
The factor of $t^2/N_\pi^3$ is combinatorial and decreases with large $N_f$ as $1/N_f$; further details are found in the Appendix. 
Increasing $N_c$ or decreasing $N_f$ enables the perturbative limit on $m_\pi/f_\pi$ to be reached for higher values of $m_\pi$.
For example, for the simplest case of $N_c=N_f=2$, dark matter mass of $m_\pi\lsim 300$~MeV can be reached; for {\it e.g.} $N_c=8$ and $N_f=2$ chiral perturbation theory is expected to break down at $m_\pi\sim 800$~MeV.

There are also 4-pion interactions induced by the kinetic and mass terms for the fermions, described via Eqs.~\eqref{eq:kin} and~\eqref{eq:eff}.
These contribute to the self-scattering cross section of dark matter, $\sigma_{\rm scatter}$, which is constrained by bullet-cluster~\cite{Clowe:2003tk,Markevitch:2003at,Randall:2007ph} and halo shape~\cite{Rocha:2012jg,Peter:2012jh} constraints to obey
\beq\label{eq:bullet}
\frac{\sigma_{\rm scatter}}{m_{\rm DM}} \lesssim 1 \units{cm^2/g}\,,
\eeq
with $m_{\rm DM}$ the dark matter mass. The self-scattering cross section, obtained along the solution to the Boltzmann equation, is plotted in the dashed curves of the left panel of Fig.~\ref{fig:3to2sol} for various values of $N_c$ and $N_f$ for an Sp($N_c$) gauge group. The results are similar for the other gauge groups and are found in Fig.~\ref{fig:others} in the Appendix. The experimental constraint is depicted by the horizontal dashed line, and provides a rough lower bound on the mass of the pions. This lower bound decreases as $N_c$ increases and $N_f$ decreases. The reason is that the self-interaction scattering cross section scales as
\beq\label{eq:xsec22}
\sigma_{\rm scatter}=\frac{m_\pi^2}{32 \pi f_\pi^4}\; \frac{a^2}{N_\pi^2}\,,
\eeq
with $a^2/N_\pi^2$ nearly constant as $N_f$ varies (see the Appendix for further details). 
For a given mass $m_\pi$, as $N_c$ increases or $N_f$ decreases, a larger value of $f_\pi$ solves the Boltzmann equation, which helps suppress the self-scattering cross section below its constrained value. For example, for the  minimal case of $N_c=N_f=2$, structure formation dictates pion masses above $m_\pi\gsim 300$~MeV; for $N_c=8$ and $N_f=2$ the lower bound is $m_\pi\gsim150$~MeV.

Combined with the upper bound from chiral perturbation theory, a rough range for the mass of the pions is obtained. For example, as depicted in the left panel of Fig.~\ref{fig:3to2sol}, the minimal case of two flavors in an Sp(2) $\simeq$~SU(2) gauge group points to pion dark matter mass of order $\sim 300$~MeV; for $N_c=8$ with $N_f=2$, the range is widened to $\sim 150-800$~MeV. The minimal case of $N_f=3$ in SU(3) and O(3) gauge groups exhibits a tension between the rough upper bound on $m_\pi$ from perturbativity and the rough lower bound on $m_\pi$ stemming from self-scattering constraints, and so are not depicted in the left panel of Fig.~\ref{fig:others}.

A comment is in order regarding higher-derivative corrections. Throughout we have used the 4-point interaction terms stemming from the mass and kinetic terms, Eqs.~\eqref{eq:kin} and~\eqref{eq:eff}. As is evident, the theory is pushed to the strongly interacting regime where $m_\pi$ is not far from the effective cutoff, $\Lambda = 2\pi f_\pi$; here higher-derivative terms may induce ${\cal O}(1)$ effects, shifting the lower bound on the pion mass accordingly. The self-scattering cross section of Eq.~\eqref{eq:xsec22} is thus a proxy, which suffices for the purpose of obtaining a characteristic  pion mass range.

Modifications to the presented canonical realization of the SIMP mechanism are possible.
For instance, it is possible to write a mass term for the confining fermions that explicitly breaks the
flavor symmetry of Sp($2N_f$), SU($N_f$) or SO($N_f$) in the class of Sp($N_c$), SU($N_c$) or O($N_c$) gauge theories.
If one pion is lighter than the others, this pion will be the dark matter. Since the WZW term, Eq.~\eqref{eq:wzw}, induces $3\to2$ interactions between five different flavors of pions,
the decay of the other pions to the lightest one must occur after freeze-out, and their masses must be close. Considering the 4-pion interactions, there are no self-interaction terms between pions of the same flavor originating from the kinetic term. In contrast, the fermion mass term of Eq.~\eqref{eq:pimass} does induce same-flavor self-scattering for the lightest pion. The resulting self-scattering cross section for the dark matter state, $\sigma_{\rm scatter}'$, is suppressed numerically between a factor of a few to an order of magnitude, depending on the gauge group, compared to the degenerate-pion case. Further details are given in the Appendix.
The rough lower bound on the mass of the dark matter is then reduced compared to the degenerate-pion scenario, expanding the allowed dark matter mass window towards lower masses.

The results for an Sp($N_c$) gauge group with a broken flavor symmetry are depicted in the right panel of Fig.~\ref{fig:3to2sol}, and the results for the SU($N_c$) and O($N_c$) gauge groups are depicted in the right panel of Fig.~\ref{fig:others} in the Appendix, for various values of $N_c$ and $N_f$. 
For instance, in the simplest case of an Sp(2)~$\simeq$~SU(2) gauge group with 2 flavors, explicit breaking of the Sp(4) flavor symmetry relaxes the self-scattering cross section constraint by an order of magnitude, such that pion masses in the range $\sim 70-300$~MeV are allowed. Similarly, with a broken flavor symmetry, the QCD-like case of an SU(3) gauge group with 3 flavors is now viable and points to pion masses of order $m_\pi\sim 150-350$~MeV.

\section{Discussion}

The two basic features of the SIMP setup --- strong $3\to2$ interactions within the dark sector and thermal equilibrium between the dark and visible sectors --- dictate observable signals for this mechanism.

The strong interactions in the dark sector give an unavoidable contribution to a $2\to2$ self-scattering cross section amongst the pions, which is constrained \`a la  Eq.~\eqref{eq:bullet}. The failure of N-body simulation to reproduce the small scale structure of galactic halos has led to the `core vs. cusp' and `too big to fail' puzzles (see {\it e.g.}~\cite{Spergel:1999mh,deBlok:2009sp,BoylanKolchin:2011de} for discussion and references). These motivate self-interacting dark matter with a scattering cross section of~\cite{ Vogelsberger:2012ku,Zavala:2012us,Rocha:2012jg,Peter:2012jh}
\beq
\left(\frac{\sigma_{\rm scatter}}{m_{\rm DM}}\right)_{\rm obs}=(0.1-10)\ {\rm cm}^2/{\rm g}\,.
\eeq
As is evident from the dashed curves in Fig.~\ref{fig:3to2sol} (and Fig.~\ref{fig:others} in the Appendix),
the simplest realization of the dark sector presented in this letter automatically yields a contribution of the right size to this cross section. Such behavior was anticipated in Ref.~\cite{Hochberg:2014dra}, though in the absence of explicit realizations of the strongly coupled sector, only a qualitative statement could be made.
The explicit realization of the dark sector presented in this letter now proves this statement quantitatively.
Altered predictions for structure formation are a signal of the SIMP mechanism.

In addition, the required thermal equilibrium between the visible and dark sectors dictates non-negligible interactions between the two. As a result, observable signals are predicted in direct and indirect detection, colliders and cosmology. In contrast to structure formation discussed above, here the precise signatures depend on the mediation mechanism between the visible and dark sectors, and will be explored in detail in future work~\cite{future}.

We find it intriguing that the resulting mass scales indicated by the SIMP mechanism are surprisingly close to the QCD scale. This suggests a possible joint dynamical origin for both the hidden and visible strong scales (see {\it e.g.}~\cite{Bai:2013xga}). Complete SIMP models of this kind will be presented in an upcoming publication~\cite{future}.

\mysections{Acknowledgments}
The work of YH is supported by the U.S. National Science Foundation under Grant No. PHY-1002399. YH is an Awardee of the Weizmann Institute of Science - National Postdoctoral Award Program for Advancing Women in Science. EK is supported by the NSF under Grant No. PHY-1316222.
HM was supported by the U.S. DOE under Contract DE-AC02-05CH11231, and
by the NSF under grants PHY-1002399 and PHY-1316783.  HM was also
supported by the JSPS Grant-in-Aid for Scientific Research (C)
(No.~26400241), Scientific Research on Innovative Areas
(No.~26105507), and by WPI, MEXT, Japan.
TV is supported by the US-Israel Binational Science Foundation, the EU-FP7 Marie Curie, CIG fellowship and by the I-CORE Program of the Planning Budgeting Committee and the Israel Science Foundation (Grant No. 1937/12). JGW is supported in part by the NSF under Grant No. PHY-0756174. This work was supported in part by the NSF under Grant No. PHYS-1066293 and the hospitality of the Aspen Center for Physics.

\section{Appendix}\label{sec:app}

\subsection{SU($2 N_f$)/Sp($2N_f$) Generators}
\begin{table*}[th!]
\begin{center}
\[
\begin{array}{c | c | c | c}
    {\bf G/H} & {\bf N_{\pi }} & {\bf t^2} & {\bf N_f^2 \left(c^2+{r^2}/{9}\right) } \\ \hline\hline
 \text{SU(2$N_f$)/Sp($2 N_f$)} & (2 N_f+1) (N_f-1) & \frac{2}{3} N_f \left(N_f^2-1\right) \left(4 N_f^2-1\right) & 4 (N_f-1) (2 N_f+1) \left(6 N_f^4-7 N_f^3-N_f^2+3 N_f+3\right) \\
 \text{SU($N_f$)$\times$SU($N_f$)/SU($N_f$)} & N_f^2-1 & \frac{4}{3} N_f \left(N_f^2-1\right) \left(N_f^2-4\right) & 8 (N_f-1) (N_f+1) \left(3 N_f^4-2 N_f^2+6\right) \\
 \text{SU($N_f$)/SO($N_f$)} & \frac{1}{2} (N_f+2) (N_f-1) & \frac{1}{12} N_f \left(N_f^2-1\right) \left(N_f^2-4\right) & (N_f-1) (N_f+2) \left(3 N_f^4+7 N_f^3-2 N_f^2-12 N_f+24\right) \\\hline\hline
\end{array}
\]
\end{center}
\caption{For an initial global symmetry $G$, which preserves a global symmetry $H$ after chiral symmetry breaking, with $N_f$ flavors: the number of pions $N_\pi$ (corresponding to the number of broken generators), the combinatorical factor $t^2$ entering the $3\to2$ cross section via Eq.~\eqref{eq:xsec32} and the factors entering the 4-point self-scattering interactions via Eq.~\eqref{eq:xsec22} with Eq.~\eqref{eq:asq}.
}\label{tab:sup}
\end{table*}

Here we construct the generators for the coset space of the Sp($N_c$) gauge theory with $N_f$ flavors ($2 N_f$ doublet Weyl fermions)
discussed in detail in this letter. The global symmetry of Eq.~\eqref{eq:L} is SU($2 N_f$), which upon chiral symmetry breaking leads to a bilinear
\begin{align}
\langle q_i q_j \rangle = \mu^3 J_{ij},
\end{align}
where
\begin{align}
J= \left(\begin{array}{cc} 0 & \mathbbm{1}_{N_f}\\ -\mathbbm{1}_{N_f} &0\end{array}\right)\,, \qquad J^2 = -\mathbbm{1}_{2N_f}
\end{align}
is a $2 N_f\times 2 N_f$ anti-symmetric matrix that preserves an Sp($2 N_f$) subgroup unbroken~\cite{Witten:1983tx}.

The vacuum is preserved for matrices $\alpha$ such that
\beq
\alpha J + J \alpha^T  = 0\,. \label{spn}
\eeq
The broken generators that span the SU($2 N_f$)/Sp($2 N_f$) coset space can be parameterized as the traceless hermitian matrices $\pi$ such that
\beq
\pi J - J \pi^T  = 0\,. \label{sunspn}
\eeq
The solutions to Eq.~(\ref{sunspn}) have the general form
\begin{eqnarray}
\pi = \left(\begin{array}{cc} A & -B^*\\B & A^T\end{array}\right)
\end{eqnarray}
where $A$ is a general $N_f \times N_f$ traceless hermitian matrix and $B$ is general antisymmetric $N_f \times N_f$  matrix.
We write $\pi=\pi^a T^a$ with $T^a$ the generators spanning the matrices $A$, $B$, normalized such that $\!\Tr\![ \pi \cdot \pi] = 2 \pi^a \pi^a$. For example, for $N_f=2$,
\beq
\pi = \left(
\begin{array}{cccc}
 \frac{\pi _3}{\sqrt{2}} & \frac{\pi _1-i \pi _2}{\sqrt{2}} & 0 & \frac{\pi _4-i \pi _5}{\sqrt{2}} \\
 \frac{\pi _1+i \pi _2}{\sqrt{2}} & -\frac{\pi _3}{\sqrt{2}} & -\frac{\pi _4-i \pi _5}{\sqrt{2}} & 0 \\
 0 & -\frac{\pi _4+i \pi _5}{\sqrt{2}} & \frac{\pi _3}{\sqrt{2}} & \frac{\pi _1+i \pi _2}{\sqrt{2}} \\
 \frac{\pi _4+i \pi _5}{\sqrt{2}} & 0 & \frac{\pi _1-i \pi _2}{\sqrt{2}} & -\frac{\pi _3}{\sqrt{2}} \\
\end{array}
\right)\,.
\eeq

\subsection{Cross Sections}

Here we construct the $3\to2$ and scattering cross sections for the pions discussed in this letter. The WZW term of Eq.~\eqref{eq:wzw} can be written in the form
\begin{equation}
{\cal L}_{\rm WZW} = \frac{2N_c}{15 \pi^2 f_\pi^5}\epsilon^{\mu\nu\rho\sigma} \sum_{\begin{smallmatrix}i<j<k\\ <l<m \end{smallmatrix}}{T_{ijklm} \pi_i \partial_\mu
  \pi_j \partial_\nu \pi_k \partial_\rho \pi_l \partial_\sigma \pi_m}.
\end{equation}
The thermally averaged cross section for a particular 5-pion process is
\beq
\langle \sigma v^2 \rangle_{ijk \to lm} = \frac{m^5_\pi N_c^2 T_{\{ijklm\}}^2}{96 \sqrt{5} \pi ^5 f_{\pi }^{10} x^2}\,,
\eeq
where $x = m_\pi/T$ and $\{\ldots\}$ denotes ordering, {\it i.e.} $\{1,3,6,8,2\}=1,2,3,6,8$. There are $N_\pi$ Boltzmann equations, where $N_\pi$ is the number of broken generators and is given in Table~\ref{tab:sup} for the different gauge groups presented in this work.
Summing them up leads to the Boltzmann equation for the total number density of pions, Eq.~\eqref{eq:bol}, with
\beq
\langle \sigma v^2\rangle_{3\to2} = \frac{5 \sqrt{5} N_c^2 m_\pi^5}{2 \pi^5 x^2 f_\pi^{10}} \frac{t^2}{N_\pi^3}
\eeq
where
\beq
t^2 \equiv \frac{1}{{5!^2}}{\sum_{} T_{\{ijklm\}}^2}\,,
\eeq
and its behavior in the different classes of symmetry structures considered here is given in Table~\ref{tab:sup}.

The scattering cross section of pions receives in principle contributions from the kinetic and mass terms of the confining fermions. The 4-point interaction arising from the kinetic term, Eq.~\eqref{eq:kin}, can be written as
\begin{equation}\begin{array}{rcl}
\mathcal{L}_{{\rm kin},\pi^4} =- \frac{1}{6 f_\pi^2} r_{ijkl}\pi_i \pi_j  \partial_{\mu} \pi_k \partial^{\mu} \pi_l
\end{array}\end{equation}
where $r_{ijkl} =f_{ikm}f_{jlm}$ with $f_{ijm}$ the structure constants of the full SU($2N_f$) flavor symmetry, and $(m,n)$ run over all generators of the original flavor symmetry [SU($2N_f$), SU($N_f$)$\times$SU($N_f$) or SU($N_f$), for the gauge groups Sp($N_c$), SU($N_c$) and O($N_c$), accordingly]. The mass term, Eq.~\eqref{eq:eff}, gives a 4-point interaction which can be written as
\begin{eqnarray}
 \Delta {\cal L}_{{\rm eff},\pi^4}=  \frac{ m_\pi^2}{ 12 f_\pi^2} c_{abcd}\pi^a \pi^b\pi^c \pi^d\,.
\end{eqnarray}
The resulting self-interacting scattering cross section for the pions is then given by Eq.~\eqref{eq:xsec22},
with
\beq\label{eq:asq}
a^2 &\equiv& \left(c^2+r^2/9\right)\,,
\eeq
where
\beq
r^2 &\equiv& \sum_{abcd}\left(r_{abcd}+r_{abdc}+r_{bacd}+r_{badc}\right)^2\,,
\eeq
and
\beq
c^2 &\equiv& \sum_{abcd}c_{abcd}^2\,,
\eeq
with their behavior for the different symmetry structures given in Table~\ref{tab:sup}.

In the case of a mass term that explicitly breaks the Sp($2N_f$) flavor symmetry, and assuming a single pion is the lightest state of the theory, only the mass term contributes to the cross section relevant for self-interactions, and we find
\beq
\sigma'_{\rm scatter} = \frac{m_\pi^2}{32\pi f_\pi^4}\; a'^2\,,
\eeq
where $a'^2\simeq1$ for an Sp($N_c$) gauge theory, or $a'^2\simeq4$ for SU($N_c$) or O($N_c$) gauge theories. As a result, one can gain an order of magnitude [factor of a few] suppression compared to the scattering cross section, Eq.~\eqref{eq:xsec22}, of the canonical realization with degenerate pions that preserves the Sp($2N_f$) [SU($N_f$) and SO($N_f$)] symmetry. The lower bound on the pion mass is relaxed accordingly.

\subsection{Other gauge group results}

\begin{figure*}[ht!]
\begin{center}
\includegraphics[width=0.5\textwidth]{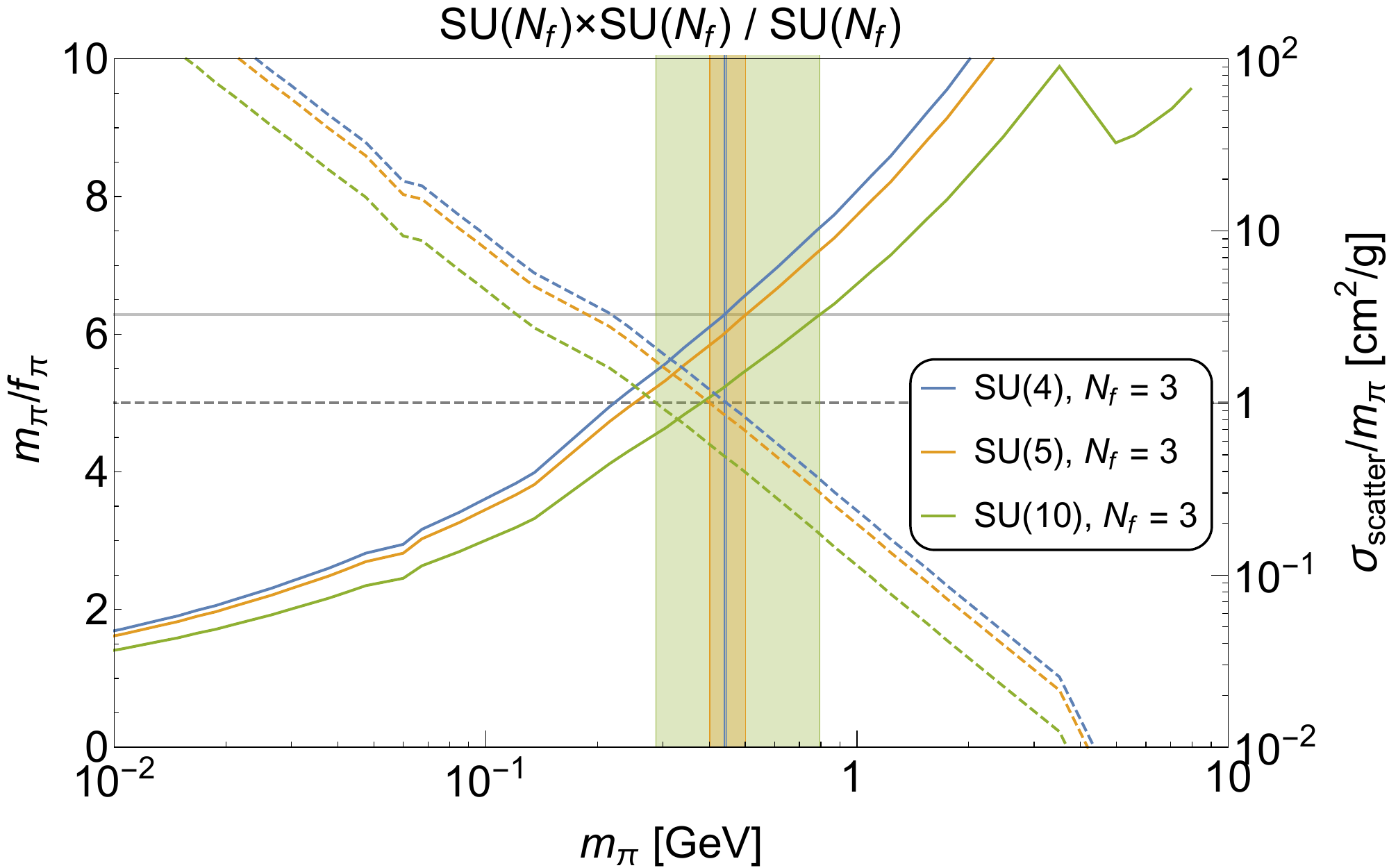}~~~~~~~\includegraphics[width=0.5\textwidth]{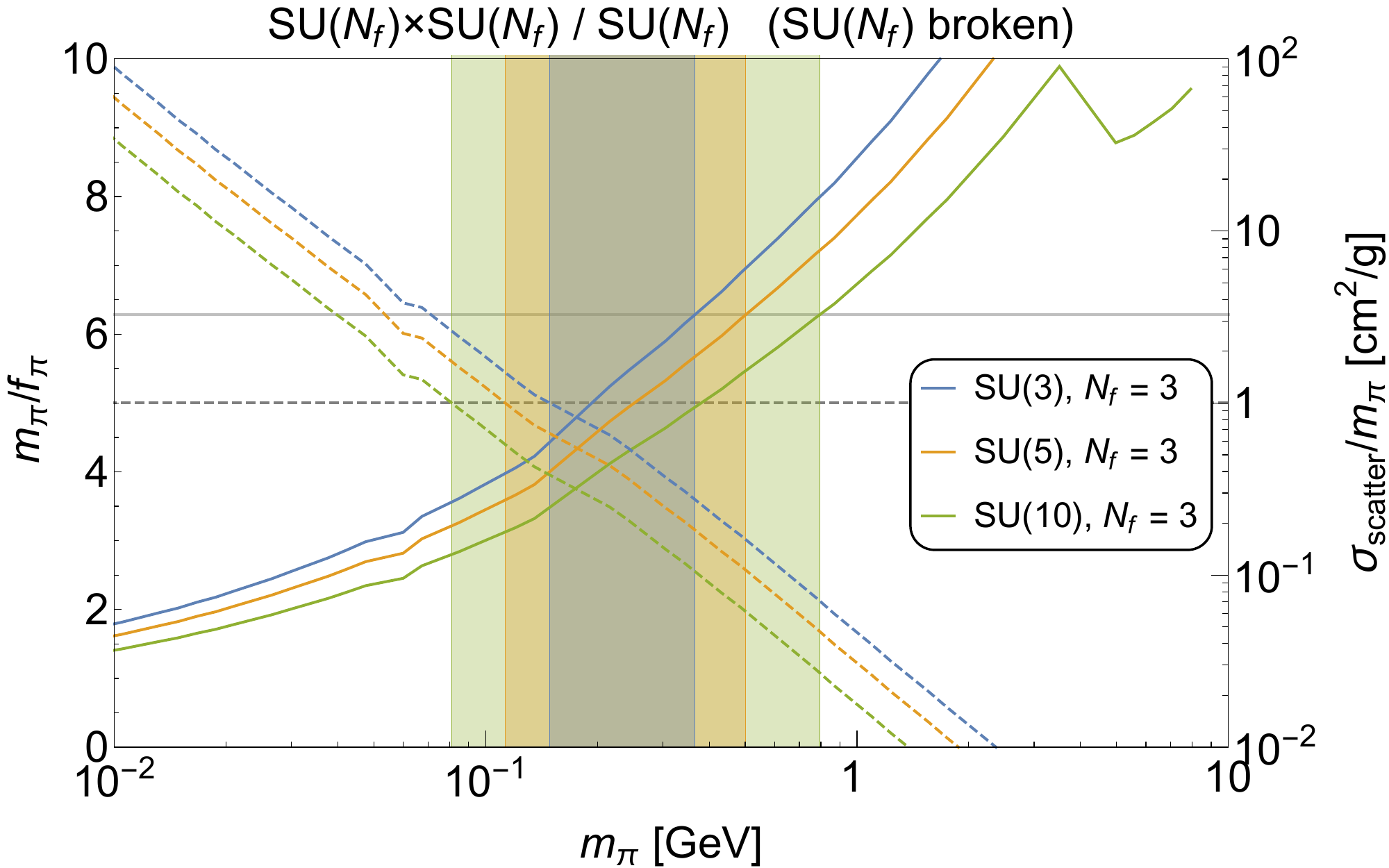}\\~\\
\includegraphics[width=0.5\textwidth]{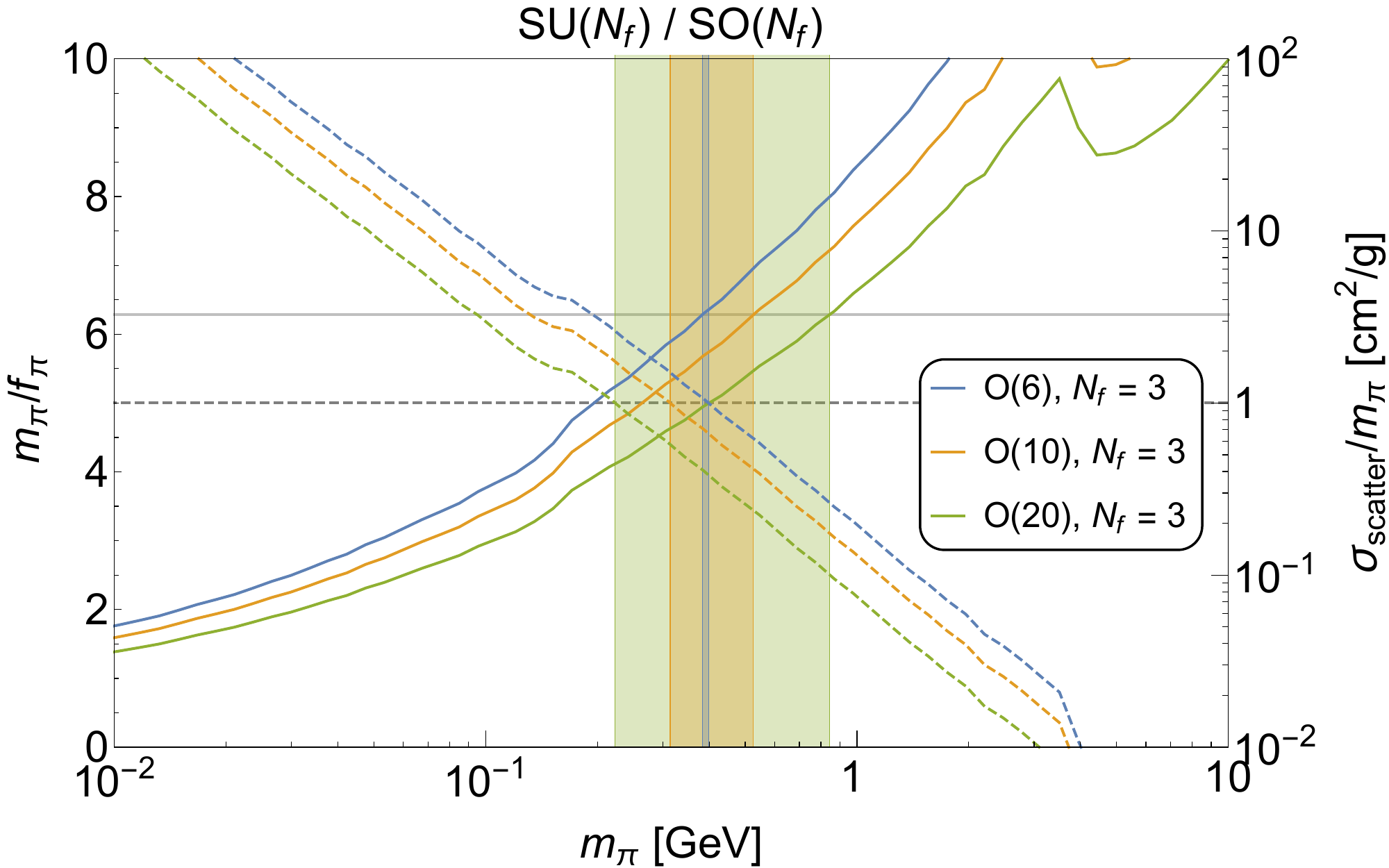}~~~~~~~\includegraphics[width=0.5\textwidth]{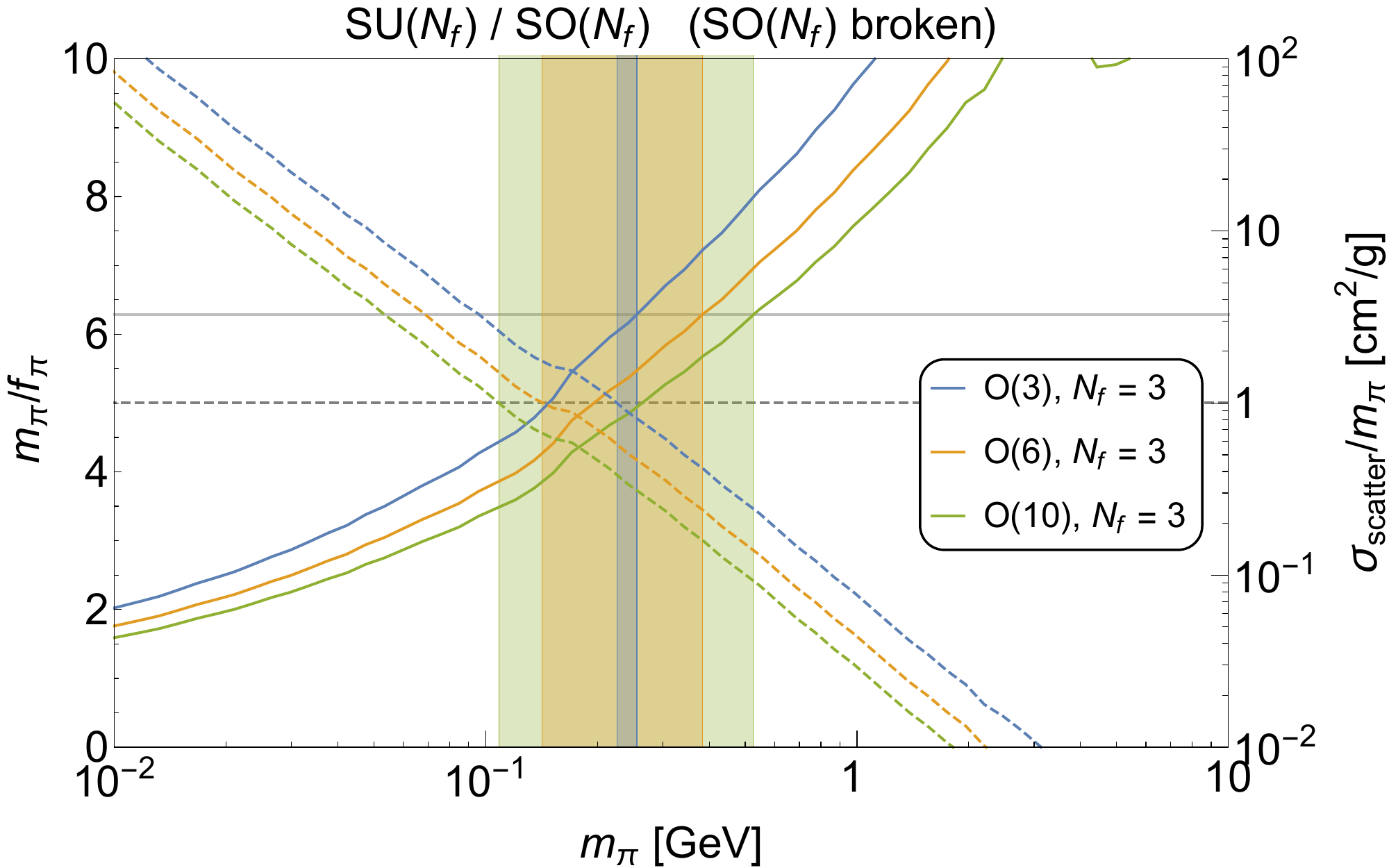}
\caption{
\label{fig:others}
{\bf Solid curves:} the solution to the Boltzmann equation of the $3\to2$ system, yielding the measured dark matter relic abundance for the pions, $m_\pi/f_\pi$ as a function of the pion mass (left axis). {\bf Dashed curves:} the self-scattering cross section along the solution to the Boltzmann equation, $\sigma_{\rm scatter}/m_\pi$ as a function of pion mass (right axis).
All curves are for selected values of $N_c$ and $N_f$, for an SU($N_c$) ({\bf top panel}) or an O($N_c$) ({\bf bottom panel}) gauge group with a conserved ({\bf left panel}) or broken ({\bf right panel}) SU($N_f$) or SO($N_f$) flavor symmetry, respectively.
The solid horizontal line depicts the perturbative limit of $m_\pi/f_\pi\lsim 2\pi$, providing a rough upper limit on the pion mass; the dashed horizontal line depicts the bullet-cluster and halo shape constraints on the self-scattering cross section, Eq.~\eqref{eq:bullet}, placing a lower limit on the pion mass. Each shaded region depicts the resulting approximate range for $m_\pi$ for the corresponding symmetry structure.
}
\end{center}
\end{figure*}
In Fig.~\ref{fig:others} we plot the results of solving the Boltzmann equation, Eq.~\eqref{eq:bol} for SU($N_c$) (top panel) and O($N_c$) (bottom panel) gauge groups, for several values of $N_c$ and $N_f$ (solid curves). The approximate perturbative limit of $m_\pi/f_\pi\lsim 2\pi$ is depicted by the horizontal solid line, placing a rough upper limit on the mass of the pions. The self-scattering cross section, obtained along the solution to the Boltzmann equation, is plotted in the dashed curves, with the constraint of Eq.~\eqref{eq:bullet} indicated via the horizontal dashed line. Each shaded region indicates the allowed range for the pion mass in this case. In the left panel, the flavor symmetry [SU($N_f$) in top panel, SO($N_f$) in bottom panel] is preserved; the right panel depicts the results when the flavor symmetry is explicitly broken. For the preserved flavor symmetry case, $N_c$ values below those depicted exhibit a tension between the perturbativity regime $m_\pi/f_\pi\lsim 2\pi$ and the self-interaction constraint of Eq.~\eqref{eq:bullet}.




%
%
%

\end{document}